\documentstyle[aps,prl,epsfig]{revtex}
\begin{document}
\draft

\title{Emergence of structural anisotropy in Optical Glasses treated
to support Second Harmonic Generation}

\author{C. Cabrillo, G.J. Cuello, P. Garc{\'\i}a-Fern{\'a}ndez and
F.J. Bermejo}
\address{Instituto de Estructura de la Materia,
Serrano 123, Madrid E-28006, Spain}

\author{V. Pruneri and P. G. Kazansky}
\address{Optoelectronics Research Centre, University of
Southampton, \\ Southampton S017 IBJ, United Kingdom}

\author{S.M. Bennington, W.S. Howells}
\address{Rutherford Appleton Laboratory, Chilton, Didcot, Oxon,
OX11 0QX, United Kingdom}

\date{\today}
\maketitle

\begin{abstract}
Structural alterations in {\it v}-SiO$_2$ induced by
"thermal poling", a treatment which makes the glass able to
double the frequency of an impinging infrared light, 
are revealed by neutron diffraction as a breakdown of the
macroscopic isotropy. This leads to concomitant changes in the 
vibrational density of states measured by inelastic neutron 
scattering. The observations are found to be consistent with 
the emergence of partial ordering within the glassy matrix 
along the direction of an electrostatic field applied during 
the poling treatment.
\end{abstract}


Few discoveries have puzzled the optics community more than
the emergence of visible (green) light from optical fibers after 
strong irradiation by an infrared laser \cite{OSTE86}. This 
frequency-doubling phenomenon known as Second Harmonic 
Generation (SHG) is not expected to take place in a centrosymmetric 
material such as the amorphous silica fiber-core, which shows no
measurable second-order optical susceptibility $\chi^{(2)} $
\cite{YARI84}. The process to be
efficient also requires well defined phase-matching between the
interacting waves to allow for constructive interference, and this
seems even more difficult to fulfill within the glassy medium.
Several plausible explanations about the origin of the phenomenon
have been put forward. One of the most widely accepted
\cite{DIAN91} does not involve structural modifications to 
accomplish the breakdown of the glass radial symmetry. Rather, it 
postulates the emergence of a spatially modulated local dc field, 
$E_{0}$, which, via a third-order nonlinearity ($\chi^{(3)}$) (finite 
in isotropic materials), induces a spatially modulated second
order nonlinearity ($\chi^{(2)} \propto \chi^{(3)} E_{0}$) able to
double the pump frequency.
The achievement of a permanent $\chi^{(2)}$ in optical glasses
has focused a large research effort, which lead to the discovery of
alternative poling techniques. In actual fact, the phenomenon can 
be produced by application of a high voltage ($\sim$ 5 kV, just 
below dielectric breakdown) to glass plates at moderate
temperatures (${\rm\sim 540 \;K - 580\; K}$, compared with
$\approx$ 1475 K where the glass melts). This method, known as 
``thermal'' poling \cite{Mye91}, provides permanent second-order 
nonlinear responses comparable to those shown by inorganic crystals. 
Whether the mechanism(s) leading to the emergence
of a second order nonlinearity in ``thermally-poled'' glasses 
differ from those of photoinduced SHG or not needs to be clarified.
At any rate, the relevant point stems from the possibility
this method has opened up for developing inexpensive
integrated optical frequency converters and electro-optic
modulators.

Up to now studies on the microscopic structure and/or
dynamical alterations associated with poling are scarce.
Significant changes between poled and native fibers at rather well
defined frequencies have been reported from Raman studies
\cite{Gab87}. Direct interpretation of such Raman data is however
hampered by the need of knowledge of the possible
structural alterations in order to assess the concomitant variations
in the Raman matrix elements governing the signal intensities.
Here we are set to investigate the microscopic mechanisms leading
to the appearance of SHG in ``thermally'' poled glasses
using neutrons as a probe. Alterations in the microscopic
structure are explored by diffraction means, whereas the
perturbations of the microscopic dynamics are addressed
by means of inelastic neutron scattering (INS) measurements of the
generalized frequency spectrum.

\begin{figure}
\centerline{\psfig{figure=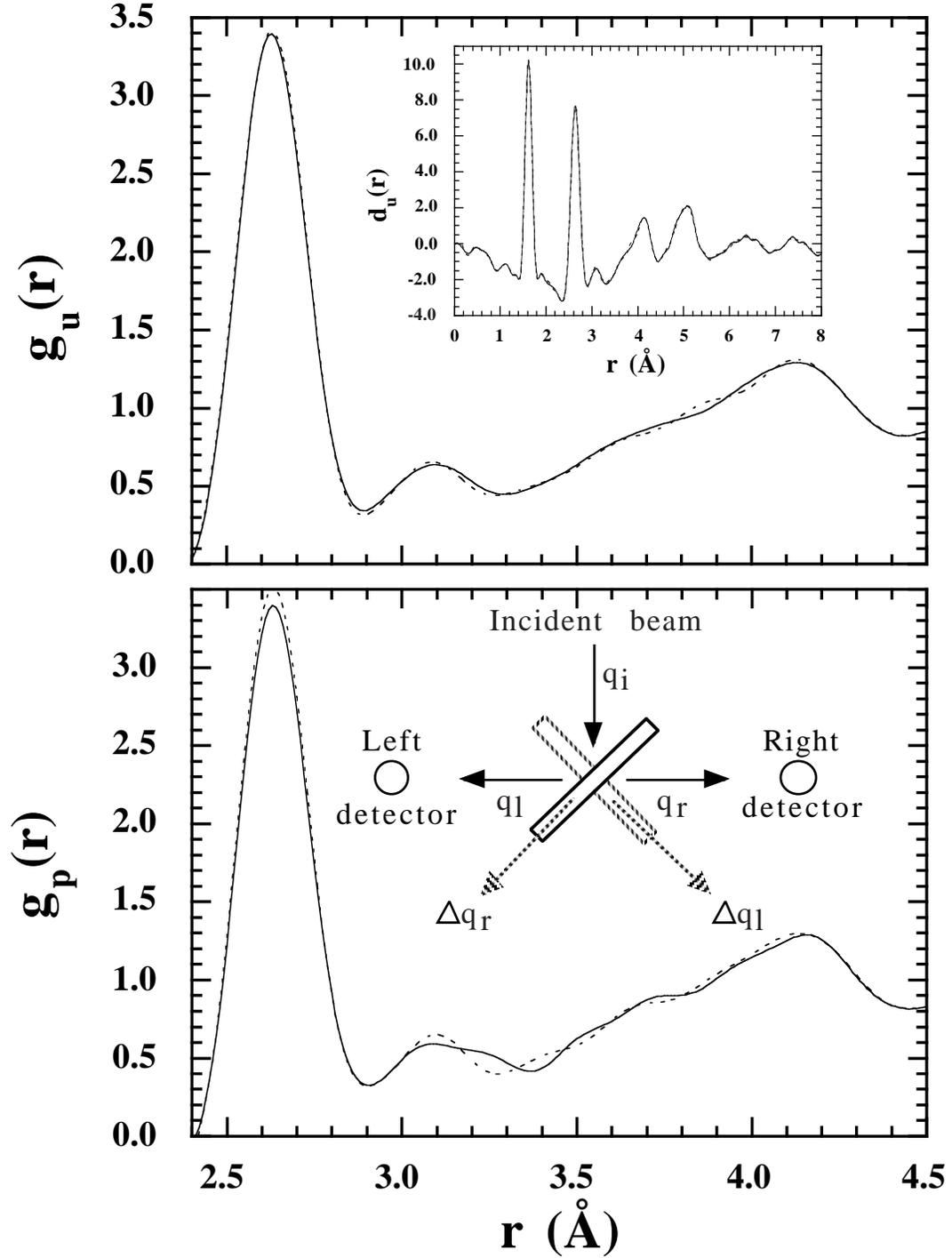}}
\caption{A comparison between the functions
$g_{\rm \parallel}(r)$ (dotted line) and $g_{\rm \perp}(r)$ (solid
line) f111 or both, the native (upper figure) and the poled samples
(lower figure). The inset in the upper figure displays the whole
measured $d_{\rm \parallel,\perp}(r)$ while the inset in the lower
figure shows the detectors/sample geometry.}
\label{statics}
\end{figure}

The samples were made of electrically-fused quartz (Infrasil,
Heraeus) with dimensions $40\times 40 \times 0.1$ mm$^3$.
The thickness ($\sim 100 \mu$m) was chosen as the smallest
possible, in order to increase the poled/unpoled volume ratio. In
fact, after treatment \cite{Mye91}, the nonlinearity is confined to a
 region $\sim 5 \mu$m deep under the anode surface regardless of
the sample thickness. Poling was carried out at 548 K under an
applied dc voltage of $\sim 5$ kV for 40 minutes.
The poled sample consisted in fifteen plates stacked together.
For comparison a similar stack of unpoled samples was used.
The measurements were performed using the LAD, time-of-flight
diffractometer and the MARI spectrometer, both at the ISIS pulsed
neutron source (R.A.L., UK). Those carried out at the diffractometer
sought to detect indications of local anisotropy induced by poling.
To see this, sets of two runs with the samples oriented
$\pm 45^{\rm o}$ with respect to the incident beam,
were performed. The diffractometer consists in an
array of 14 neutron counters covering a 2$\pi$ angular range at
discrete
intervals each one spanning a large extent of momentum-transfers.
In particular, our interest will here be focused on the analysis of
those detectors $\pm 90^{\rm o}$ with respect to the incident beam
which span momentum-transfers covering directions mostly normal
and parallel to the plates. Two spectra for both directions of
momentum-transfers are then obtained with the
corresponding detectors interchanged (see inset in Fig. \ref{statics}).
In doing so, exploration along the direction of the field applied
during poling and normal to it is envisaged. After averaging over 
both orientations, systematic errors are minimized.

The experimentally-accessible quantity of interest is here the static
pair correlation function $d_{\rm \parallel,\perp}(r)$ or
alternatively, the radial distribution function
$g_{\rm \parallel,\perp}(r)$, both related to the observable static
structure factor $S(Q)$ by a sine-Fourier transform, $d_{\rm
\parallel,\perp}(r)=
4\pi\rho r [g_{\rm \parallel,\perp}(r) -1]=\frac{2\pi}{Q}\int
{\rm d}Q Q (S_{\rm \parallel,\perp}(Q)-1) \sin(Qr)$.
The subscripts refer to directions parallel or normal to the
stack of plates and $\rho$ stands for the average number-density of the glass.
The $d_{\rm \parallel,\perp}(r)$  contains information about the 
thermally-averaged distribution of interatomic distances.
For low values of $r$ one expects to find sharp peaks corresponding
to distances separating directly bonded atoms, such as Si--O or
those O--O, which result from a pair of atoms linked to a common Si
nucleus. At larger distances the structure is progressively washed out
because of the static disorder and no measurable hints of ordering
are expected well beyond some 20~\AA.

The Fig.\ \ref{statics} depicts the $g_{\rm \parallel,\perp}(r)$
functions measured for the poled plates and compared to that
obtained for the native glass.
A glance to the curves drawn there reveals a large difference in
static correlations of the poled samples between parallel and
perpendicular directions concerning distances within the 2.8--3.4
\AA~ range. The graphs of the native samples serve to quantify
any source of systematic error.
The shortest distance, which corresponds to Si--O bonds, appears
at 1.6175 (6) \AA~ and that regarding O--O correlations shows
up at 2.641 (1) \AA. The differences between measurements mostly
parallel and normal to the native glass plates are very small
(differences in peak position of 2$\times$10$^{-3}$ \AA~ and 7
$\times$10$^{-3}$ \AA in width). In contrast, the curves for the
poled samples show that:
(a) $g_{\rm \parallel}(r)$ is basically unchanged after poling,
(differences in the peak center and linewidth are, within the error
bars indistinguishable from those of the unpoled plates),
(b) $g_{\rm \perp}(r)$ of the poled glass shows a peak at distances
characteristic of Si--Si correlations significantly wider than that
of the native sample, and
(c) clear hints of a bimodal distribution of Si--Si distances
are seen in the latter function.
In other words, the difference between $g_{\rm \parallel}(r)$ and
$g_{\rm \perp}(r)$ unequivocally shows the presence of an
exceedingly strong anisotropy at scales of about 3 \AA, which
correspond to Si--Si correlations, that is, to distances between the
centers of neighboring tetrahedra. More specifically, after a
decomposition of the
$g_{\rm \parallel,\perp}(r)$ functions below 5 \AA~into a sum of
Gaussians, the Si--Si peak appears in the native sample centered at
3.060 (9) \AA~with a width of 0.115 (8) \AA~in both $\parallel$
and $\perp$ directions.
The peak shape of the same feature in $g_{\perp}(r)$ for the poled
plates can be reproduced by addition of a further Gaussian which is
now centered at 3.228 (23) \AA~and has a width of 0.106 (8) \AA.
The ratio of the integrated intensities of these two peaks at high--
and low--$r$ is of 0.008/0.140, that is about 5.7 per cent, a figure
which becomes fairly close to the ratio of poled/unpoled material.

Some additional alterations in the glass structure are revealed in
 the difference function $\Delta d(r)=d_{\rm poled}(r)-d_{\rm
unpoled}(r)$ between poled and unpoled samples, this time $d(r)$
being evaluated from an average taken over all detectors.
There, a negative peak appearing at 4.4 \AA~ merits to be 
commented on, since this corresponds to a real-space manifestation 
of a difference in height of the first diffraction peak which shows 
its maximum at $\approx$ 1.5 \AA$^{-1}$.
The observed effects are commensurable with the increase in the 
Si--Si distance upon poling since Si--O, O--O and Si--Si correlations 
contribute to
the first peak in $S(Q)$ with an in-phase oscillation in their partial
structure factors, and therefore a mismatch of one of these 
correlations will lead to a decrease in the peak height.

The microscopic details of the structure within the poled region are
difficult to access because of the weakness of the signals and the
mechanism by which how this happens can only be a matter of
educated guessing.
However, both the value of 3.22 \AA~of the Si--Si distance within
the poled zone as well as the decrease in correlations about 4
\AA~provide indications of a structure more open than that of the
native glass which arises in the direction perpendicular to the plate
surface as a consequence of poling.
The value for the Si--O--Si bond angle would then reach some
167$^{\rm o}$, which although far larger than that of 142$^{\rm o}$
of the unpoled glass sits within the range of values found for some
polymorphs of $\rm Si O_2$ (see Table I of \cite{priya}).
On the other hand, most of the known structures of the disordered
phases of $\rm Si O_2$ show fairly wide distributions for the
referred angle, meaning that values as large as that quoted above
still have a very substantial probability of occurrence (see for
instance Fig.\ 5 of \cite{martin_d}).

The possibility of inducing a transformation from the isotropic glass
to a less disordered form can be understood on the basis of recent
results \cite{rosa}, which show how disordered crystal phases can
be formed from a glassy (or deeply supercooled) state.
In the particular case of silica polymorphs a common parent
structure corresponding to a disordered b.c.c. crystal
can be found, from where the known forms of $\rm Si
O_2$ can be derived \cite{salje} by ordering and
displacive mechanisms which involve pressure and the degree of
occupancy of the b.c.c. structure by oxygen atoms.
It seems then plausible to expect an ordering transition between
the fully disordered and some partially ordered structure occurring 
via the referred cubic crystal as a consequence of the aligning 
electric-field. Large scale particle rearrangements within the glassy
matrix can be expected to occur under the harsh conditions of 
poling. A glaring example of this is provided by the recent 
observation of full crystallization after segregation towards 
the surface of some of the components of soda-lime silicate 
glass-plates as a consequence of poling under conditions very close
 to those herein employed \cite{gaspar}. 

The resulting structure may share some 
characteristics with some of the disordered crystalline polymorphs
 of $\rm Si O_2$ such as $beta-$cristobalite since:
(a) the density of the latter is very close to that of the glass, so
that a transformation at ambient pressure can be envisaged, (b) the 
poling field is applied at temperatures where the $\beta-
$cristobalite structure becomes the stable crystalline form at such
density, and c) the rather open Si--O--Si bond angle is also 
reminiscent of that proposed time ago for an idealized structure of 
high-temperature cristobalite, a form which although dynamically 
unstable \cite{priya} may possibly exist quenched within the glass 
after the poling field is removed.

The $Z(\omega)$ frequency spectrum (vibrational density of states),
has been measured using incident energies of 220 meV and 150 meV in
order to cover kinematic ranges with different resolution in
energy-transfers. To pursue the derivation of $Z(\omega)$ from the
measured data, the
procedure described in Ref. \cite{Daw96}, was followed.
It consists in an iterative scheme designed to correct for a number
of instrumental and sample-dependent (multiphonon) effects for
coherently-scattering materials, and has
been shown to give results in good agreement with
thermodynamics \cite{Daw96}.
The same procedures were followed to derive the spectra for both
poled $Z_{\rm p}(\omega)$ and native $Z_{\rm u}(\omega)$ materials.
Therefore, the $Z_{\rm p}(\omega) -Z_{\rm u}(\omega)$ difference
function should be free of systematic errors.
Measurements along different directions as done at the
diffractometer are here far more complicated so that only
an angular average was considered.

\begin{figure}
\centerline{\psfig{figure=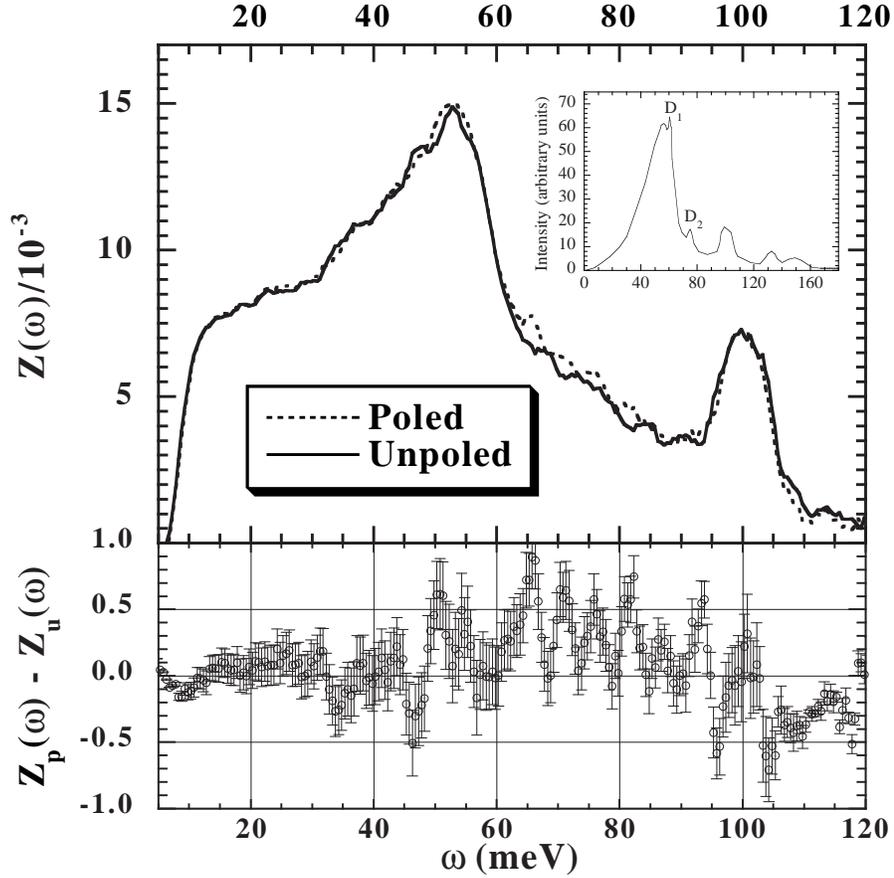,height=21cm}}
\caption{The $Z(\omega)$ frequency distribution of both, the
native (solid line) and poled  (dashed line) samples and the
difference between them. Error bars in $S(\omega)$
($\sigma_S(\omega)$)
are proportional to the square root of the number of counts in each
detector. For the DOS, the error bars are obtained taking the
semidifference between the DOS calculated with
$S(\omega)+\sigma_S(\omega)$ and $S(\omega)-
\sigma_S(\omega)$ as
input. After a smoothing process over five points, the error was
divided by \protect$5^{1/2}$. The inset depicts a schematic 
representation of the HH-polarized Raman spectrum of
bulk v-silica, showing the location of the $D_1$ and $D_2$ defect 
modes.}
\label{zeta}
\end{figure}

The Fig.\ \ref{zeta} displays a comparison between the
distributions for native and poled samples up to 120 meV ($\rm
\approx 960 \; cm^{-1}$).
The spectra show as main features  a low frequency region (limited 
by resolution effects) with a shoulder at 20 meV, a maximum at 
about 53 meV and a narrow peak at 100 meV.
Additional structure appears at higher frequencies, such as the
well known double peak with maxima at 135 meV ($\rm \approx
1080 \; cm^{-1}$) and 150 meV ($\rm \approx 1200 \; cm^{-1}$) meV.
The frequencies explored here comprise those
where previous Raman studies found changes induced by poling.
In spectra shown in Fig.\ \ref{zeta}, a measurable difference
between poled and unpoled glasses is seen for a band comprising 
60 meV ($\rm \approx 480 \; cm^{-1}$) $\leq \omega \leq$ 95 
meV 
($\rm \approx 760 \; cm^{-1}$)
and for the shape of the peak at 100 meV ($\rm \approx 800 \;
cm^{-1}$).
The differences, especially on the band about 60 meV have to be
regarded as rather large, account made of the fact that the strong
effect expected on the basis of the large anisotropy seen by
diffraction is here averaged.
The observed changes comprise those seen in previous Raman
studies \cite{Gab87}, and therefore, our
results indicate that the alterations, as seen in the dynamics, are
far more widespread than previously reckoned.
The measurements carried out using a larger incident energy (220
meV) confirmed the above mentioned changes and enabled the
exploration of the spectral region extending up to 180 meV ($\rm
\approx 1440 \; cm^{-1}$) which comprises the strong double-peak
structure referred above \cite{Car85_Ara91}.
No significant changes upon poling were found within that range of
frequencies.

The assignment of features in $Z(\omega)$ to microscopic ``modes''
still is a matter of debate.
A rationalization could be pursued in terms of models
developed to describe the high-frequency Raman spectrum
\cite{THOR82}.
Accordingly, the broad peak about 53 meV would correspond to the
so called $\omega_1$ mode, where the oxygens undergo a
symmetric stretch whereas the Si atoms stand still.
That appearing at $\approx$ 100 meV is assimilated with
the $\omega_3$ mode, where all atoms move.
The model is supplemented by two ``defect'' modes, introduced to
explain the origin of the two narrow  D$_1$ and D$_2$ Raman lines
appearing at $\approx$ 61 meV (495 cm$^{-1}$) and $\approx$ 75
meV (606 cm$^{-1}$) (see inset of Fig.\ \ref{zeta}).
Their origin is sometimes ascribed to the presence of
intermediate-range-order. A schematic drawing of the polarized
(HH) Raman response from vitreous silica is shown in the inset of
Fig.\ \ref{zeta} for comparison purposes. From the diffraction
results, one then expects changes in the
frequency ranges associated with variations in intermediate-range
order because the latter has been altered.
This is indeed the case since the D$_1$ and D$_2$ lines are within
the region where clear differences between poled and unpoled 
spectra
are seen. On the other hand, the narrowing of the 100 meV peak
can be plausibly attributed to a reduction in the spread of
inter-tetrahedral angle \cite{Gal91}, for which some hints are
also provided by the diffraction results.

On the basis of the previous discussion about the structural
modification, a comparison with results regarding the spectra of
highly disordered silica polymorphs \cite{martin_d}, seems
adequate. That concerning the strongly disordered cubic phase of
$\beta-$cristobalite shows a broad maximum at about 62 meV
\cite{martin_d}. The increase in intensity within the 65 meV
$\leq\omega\leq$ 90 meV frequency interval, seen upon poling
would then be compatible with the emergence within the poled
region of a structure somewhat reminiscent of that for the alluded 
polymorph.

In summary, large microscopic alterations are found in fused
silica after the material is subjected to ``thermal poling''.
These result in the breakdown of isotropy occurring within the
glass which involves at least next-nearest neighbors.
Such directional ordering along the direction of the poling field has, 
as a 
direct dynamical consequence, the alterations comprising a range of 
frequencies between 65 meV and 100 meV seen in the vibrational 
density of states. Whether such alterations can account for the 
measured optical second order nonlinearity needs to be clarified 
possibly by computational means for which the collected
data constitutes an invaluable benchmark.

\acknowledgments
Work supported in part by grants No. TIC95-0563-C05-03, No.
PB96-00819, CICYT, Spain, and Comunidad de Madrid 06T/039/96.
V. Pruneri acknowledge Pirelli Cavi (Italy) for
his fellowship.

\end{document}